\documentclass[final,3p,times,twocolumn,authoryear,longtitle]{elsarticle}

\usepackage{amssymb}
\usepackage{amsmath}
\usepackage{booktabs,caption}
\usepackage{threeparttable} %[flushleft]
\usepackage[T1]{fontenc}
\usepackage[utf8]{inputenc}
\usepackage{color} %in order to have colored text
\usepackage{hyperref}

\usepackage{aas_macros}

\usepackage{lineno}

\newcommand{\td}[1]{\, \mathrm{d} #1 \,}

\newcommand{\est}[3]{\left( \frac{#1}{#2} \right)^{#3}}
\newcommand{\g}{\ensuremath{\gamma}}
\newcommand{\p}{^{\prime}}

\newcommand{\E}[1]{\times 10^{#1}}

\newcommand{\onehale}{\texttt{OneHaLe}}
\newcommand{\source}{{PKS~0903$-$57}}
\newcommand{\hess}{H.E.S.S.}
\newcommand{\fermi}{\textit{Fermi}-LAT}
\newcommand{\stat}{_{\rm stat}}
\newcommand{\sys}{_{\rm sys}}
\newcommand{\thestar}{\textit{the star}}
\newcommand{\CD}{C\hspace{-0.06667em}D}
\newcommand{\nF}{\nu\hspace{-0.06667em}F_{\rm obs}}
\newcommand{\Eto}{E_{t,{\rm obs}}}
\newcommand{\Ego}{E_{\gamma,{\rm obs}}}
\journal{Journal of High Energy Astrophysics}

\begin{document}

\begin{frontmatter}



\title{Scrutinizing the 2020 multiwavelength outburst of \source\ through observations with \hess} %% Article title


%\author[LSW]{M.~Zacharias\corref{cor}} %% Author name
%\ead{contact.hess@hess-experiment.eu}

%\affiliation[LSW]{organization={Landessternwarte, Universit\"at Heidelberg},%Department and Organization
%            addressline={K\"onigstuhl}, 
%            postcode={D 69117}, 
%            postcodesep={},
%            city={Heidelberg},
%            %state={},
%            country={Germany}}

%%%%%%%%%%%%%%%%%%%%%%%%%%%%%%%%%%%%%%%%%
\author[USD]{A.~Acharyya} % HESS member, orcid 0000-0002-2028-9230
\author[DIAS,MPIK]{F.~Aharonian} % HESS member, orcid 0000-0003-1157-3915
\author[LSW]{F.~Ait~Benkhali}
\author[IRFU]{A.~Alkan}
\author[LLR]{H.~Ashkar}
\author[UNAM,NWU]{M.~Backes} % HESS member, orcid 0000-0002-9326-6400
\author[DESY]{V.~Barbosa~Martins}
\author[UP]{R.~Batzofin} % HESS member, orcid 0000-0002-5797-3386
\author[APC]{Y.~Becherini} % HESS member, orcid 0000-0002-2115-2930
\author[DESY,HUB]{D.~Berge} % HESS member, orcid 0000-0002-2918-1824
% ...

\affiliation[USD]{organization={University of Southern Denmark},
    addressline={Campusvej 55, 5230 Odense M},
    %postcode{},
    %postcodesep{},
    %city{},
    country={Denmark}}
\affiliation[DIAS]{organization={Astronomy \& Astrophysics Section, School of Cosmic Physics, Dublin Institute for Advanced Studies},
    addressline={DIAS Dunsink Observatory, Dublin D15 XR2R},
    %postcode{},
    %postcodesep{},
    %city{},
    country={Ireland}}
\affiliation[MPIK]{organization={Max-Planck-Institut für Kernphysik},
    addressline={P.O. Box 103980, D 69029 Heidelberg},
    %postcode{},
    %postcodesep{},
    %city{},
    country={Germany}}
\affiliation[LSW]{organization={Landessternwarte, Universität Heidelberg},
    addressline={Königstuhl, D 69117 Heidelberg},
    %postcode{},
    %postcodesep{},
    %city{},
    country={Germany}}
\affiliation[IRFU]{organization={IRFU, CEA, Université Paris-Saclay},
    addressline={F-91191 Gif-sur-Yvette},
    %postcode{},
    %postcodesep{},
    %city{},
    country={France}}
\affiliation[LLR]{organization={Laboratoire Leprince-Ringuet, École Polytechnique, CNRS, Institut Polytechnique de Paris},
    addressline={F-91128 Palaiseau},
    %postcode{},
    %postcodesep{},
    %city{},
    country={France}}
\affiliation[UNAM]{organization={University of Namibia, Department of Physics},
    addressline={Private Bag 13301, Windhoek 10005},
    %postcode{},
    %postcodesep{},
    %city{},
    country={Namibia}}
\affiliation[NWU]{organization={Centre for Space Research, North-West University},
    addressline={Potchefstroom 2520},
    %postcode{},
    %postcodesep{},
    %city{},
    country={South Africa}}
\affiliation[DESY]{organization={Deutsches Elektronen-Synchrotron DESY},
    addressline={Platanenallee 6, 15738 Zeuthen},
    %postcode{},
    %postcodesep{},
    %city{},
    country={Germany}}
\affiliation[UP]{organization={Institut für Physik und Astronomie, Universität Potsdam},
    addressline={Karl-Liebknecht-Strasse 24/25, D 14476 Potsdam},
    %postcode{},
    %postcodesep{},
    %city{},
    country={Germany}}
\affiliation[APC]{organization={Université Paris Cité, CNRS, Astroparticule et Cosmologie},
    addressline={F-75013 Paris},
    %postcode{},
    %postcodesep{},
    %city{},
    country={France}}
\affiliation[HUB]{organization={Institut für Physik, Humboldt-Universität zu Berlin},
    addressline={Newtonstr. 15, D 12489 Berlin},
    %postcode{},
    %postcodesep{},
    %city{},
    country={Germany}}
% ...
%%%%%%%%%%%%%%%%%%%%%%%%%%%%%%%%%%%%%%%%%

\cortext[cor]{Corresponding Author\\\textit{Email address:} \href{contact.hess@hess-experiment.eu}{contact.hess@hess-experiment.eu}}

%\newpageafter{author}

%% Abstract
\begin{abstract}
%\comment{250 word limit!}
The blazar \source\ has recently been classified as a flat spectrum radio quasar at a redshift of $z=0.2621$. In March and April 2020, \fermi\ and AGILE reported tremendous activity in high-energy \g\ rays with the flux increasing by $\sim$2 orders of magnitude compared to quiescence. The flare was observed with \hess\ in very-high-energy \g\ rays for six nights with a total observation time of 13.1\,h, resulting in the discovery of \source\ in this energy band with 
an average flux of $1.5\E{-10}\,$ph\,cm$^{-2}$s$^{-1}$ above an energy threshold of $\sim 180\,$GeV corresponding to $60\%$ of the Crab Nebula flux above the same threshold. The very-high-energy \g-ray flux was strongly variable.
X-ray and optical data were collected with \textit{Swift} and ATOM, and also indicate significant variability. The observed multiwavelength flux and spectral variability during the \hess\ observation window suggest variability time scales on the order of a few hours and reveal complex correlation patterns. 
The lack of absorption beyond that of the extragalactic background light in the \g-ray domain suggests that the emission region was located outside of the broad-line region. A leptonic one-zone modeling of the six \hess\ observation nights using the dusty torus as seed photons for the inverse-Compton scattering, results in a low magnetization of the emission region. This implies that shock acceleration is likely the main driver during the event.
\end{abstract}



%% Keywords

\end{frontmatter}

%\linenumbers


%
%%%%%%%%%%%%%%%%%%%%%%%%%%%%%%%%%%%%%%%%%%%%%%%%%%%%%%%%%%%%%%%%%%%%%%%%%%%%%%%%%%%%%%%%%
%________________________________________________________________________________________
%

\section{Introduction} \label{sec:intro}
%
% SP (summary of the Goldoni+ results), MZ
%
\source\ is a blazar, meaning an active galactic nucleus (AGN) with the jet aligned closely to the line of sight to Earth \citep{blandfordrees74}. Blazars are classified 
into flat spectrum radio quasars (FSRQs) and BL Lac objects according to the presence of broad spectral emission lines (with the split at equivalent width $EW=5\,$\AA). 
The spectral energy distribution (SED) of blazars is characterized by two broad components of which the low-energy one is due to synchrotron emission of relativistic electrons. The high-energy component is commonly attributed to inverse-Compton (IC) scattering of soft photon fields by the same electron population. The soft photon fields may be the synchrotron photons produced by the electrons themselves (synchrotron-self Compton, SSC), or from external sources such as the accretion disk (AD), the broad-line region (BLR) or the dusty torus (DT). Another possibility for the high-energy component is interactions of relativistic protons with the ambient magnetic and photon fields, which however will not be studied here \citep[see][for a review of the various radiation processes]{cerruti20}.

The broadness of the spectral components requires a similarly broad energy distribution of the underlying particles. The required acceleration is typically invoked to take place at shocks \citep{marschergear85}, through magnetic reconnection \citep{giannios+09} or other turbulent scenarios such as shear acceleration \citep[e.g.][]{riegerduffy22,wang+24}. Recent observations of X-ray polarization with IXPE suggest acceleration at shocks as the predominant mechanism in BL Lac objects \citep{liodakis+ixpe22}. The acceleration process plays a pivotal role in explaining the observed strong multiwavelength (MWL) variability occuring on various time scales. In turn, the observed variability patterns can be used to infer the underlying acceleration process through explicit modeling as summarized by, e.g., \cite{boettcher19}.

\source\ was first detected in the radio band with the Parkes telescope and included in its initial catalog of radio sources \citep{Bolton+1964}. Its position is well established 
(RA$_{\rm J2000}$: 09$^h$ 04$^m$ 53.18$^s$; DEC$_{\rm J2000}$: $-$57$^{\circ}$ 35$\p$ 05.78$^{\prime\prime}$), making it one of the reference sources for the International Celestial Reference Frame \citep[ICRF,][]{Charlot+2020}. The optical counterpart was discovered by \cite{White+1987}. Observations with GAIA have been used to resolve this counterpart into two sources \citep{gaia23}: one at the position of the radio source with a magnitude $G=17.9$, and a brighter Galactic star (hereafter \thestar) with a magnitude $G=15.9$ located only 0.67'' away. X-ray emission is also detected at the position of the radio source. A one-sided arcsecond-scale jet to the northeast of the core is observed in both radio and X-rays \citep{Marshall+2005}, confirming the AGN nature of the source. Additionally, high-energy (HE, $100\,$MeV$<E<100\,$GeV) \g-ray emission has been detected with \fermi\ \citep{abdollahi22}. 

Confusion prevailed until recently regarding the nature and redshift of \source. The source is classified as a blazar candidate of uncertain type in the \fermi\ 12-year source catalogue \citep[4FGL-DR3,][]{abdollahi22}. The only reported measurement of its redshift was by \cite{Thompson+1990} based on a low signal-to-noise optical spectrum displaying weak features at 4751\,\AA\ and 6306\,\AA\ superimposed on a power-law component. These features were identified as MgII and [OII], suggesting a redshift of $z=0.695$. However, the source identified by \cite{Thompson+1990} is separated by 4’’ from the well-known radio position of \source, indicating a potential error in counterpart identification. The study of the optical spectrum of \source\ has been challenging due to the presence of \thestar. Recently, \cite{Goldoni+24} effectively separated the two sources using observations performed with FORS2 at the VLT with excellent seeing ($\sim0.5$’’), providing a robust measurement of the source's redshift at $z=0.2621\pm0.0006$. They also suggest an FSRQ classification for \source, based on the detection of a broad H$\alpha$ emission line with an $EW$ of 8\,\AA.

In March and April 2020, \source\ underwent a strong HE \g-ray outburst as reported by the AGILE and \fermi\ teams \citep{lucarelli+20,buson20}. The HE flux varied strongly on time scales of a few hours \citep{Shah+21,Mondal+21}. The announcements of the flare triggered follow-up observations with the High Energy Stereoscopic System (\hess) in the very-high-energy (VHE, $E>100\,$GeV) \g-ray band and other MWL facilities. 
The analysis of the \hess\ and MWL data (from \fermi, \textit{Swift}, and ATOM) is presented in Sec.~\ref{sec:data}. The study of flux and spectral variability, as well as a leptonic modeling of the data recorded around the \hess\ observations are described in Secs.~\ref{sec:varia} and~\ref{sec:model}, respectively. The results are summarized and discussed in Sec.~\ref{sec:discon}.

%
%%%%%%%%%%%%%%%%%%%%%%%%%%%%%%%%%%%%%%%%%%%%%%%%%%%%%%%%%%%%%%%%%%%%%%%%%%%%%%%%%%%%%%%%%
\section{Data Analysis} \label{sec:data}
\begin{figure}
\centering
\includegraphics[width=0.48\textwidth]{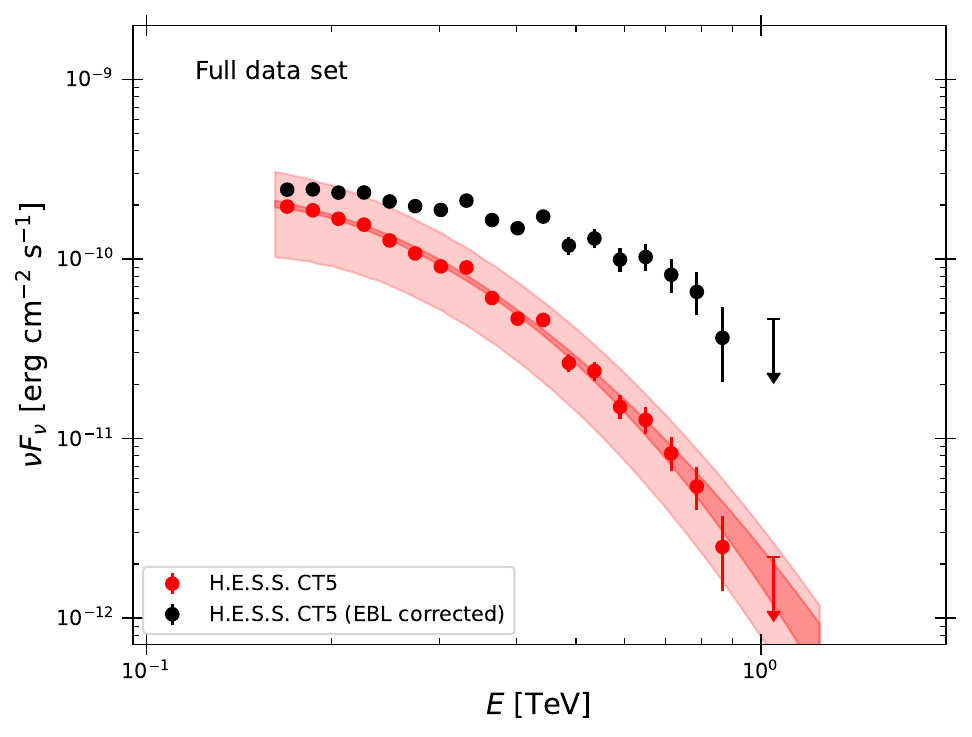}
\caption{
Total observed VHE \g-ray spectrum from H.E.S.S. observations with CT5 (red). The black points are corrected for EBL absorption using the model by \cite{Franceschini+08} at $z=0.262$. The dark and light butterflies display statistical and systematic errors, respectively, while error bars are statistical only. Upper limits are given at $95\%$ confidence level.
}
\label{fig:spec_hess_full}
\end{figure}
\begin{figure*}
\centering
\includegraphics[width=0.98\textwidth]{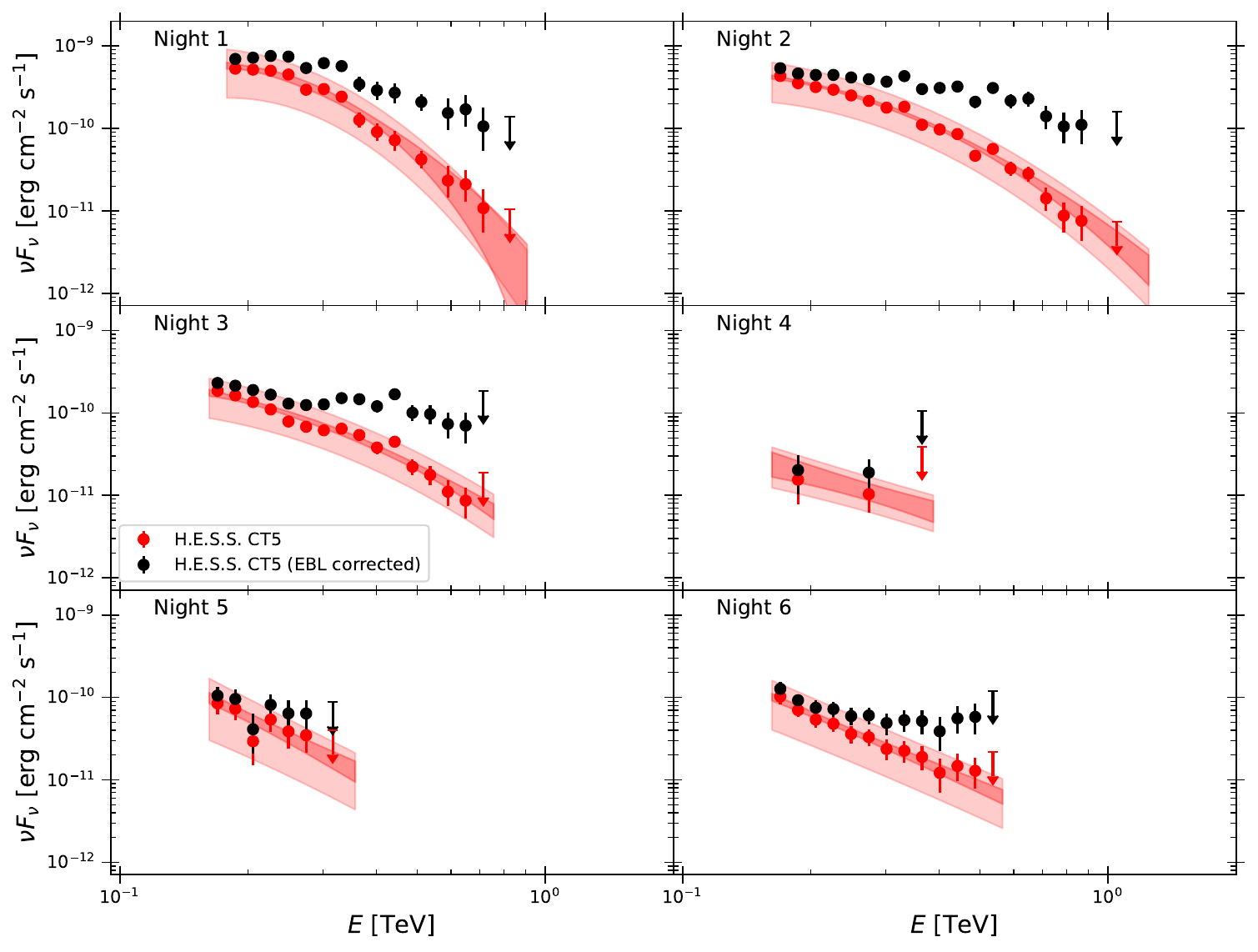}
\caption{Same as Fig.~\ref{fig:spec_hess_full}, but for individual nights. 
}
\label{fig:spec_hess_night}
\end{figure*}
\begin{figure*}
\centering
\includegraphics[width=0.98\textwidth]{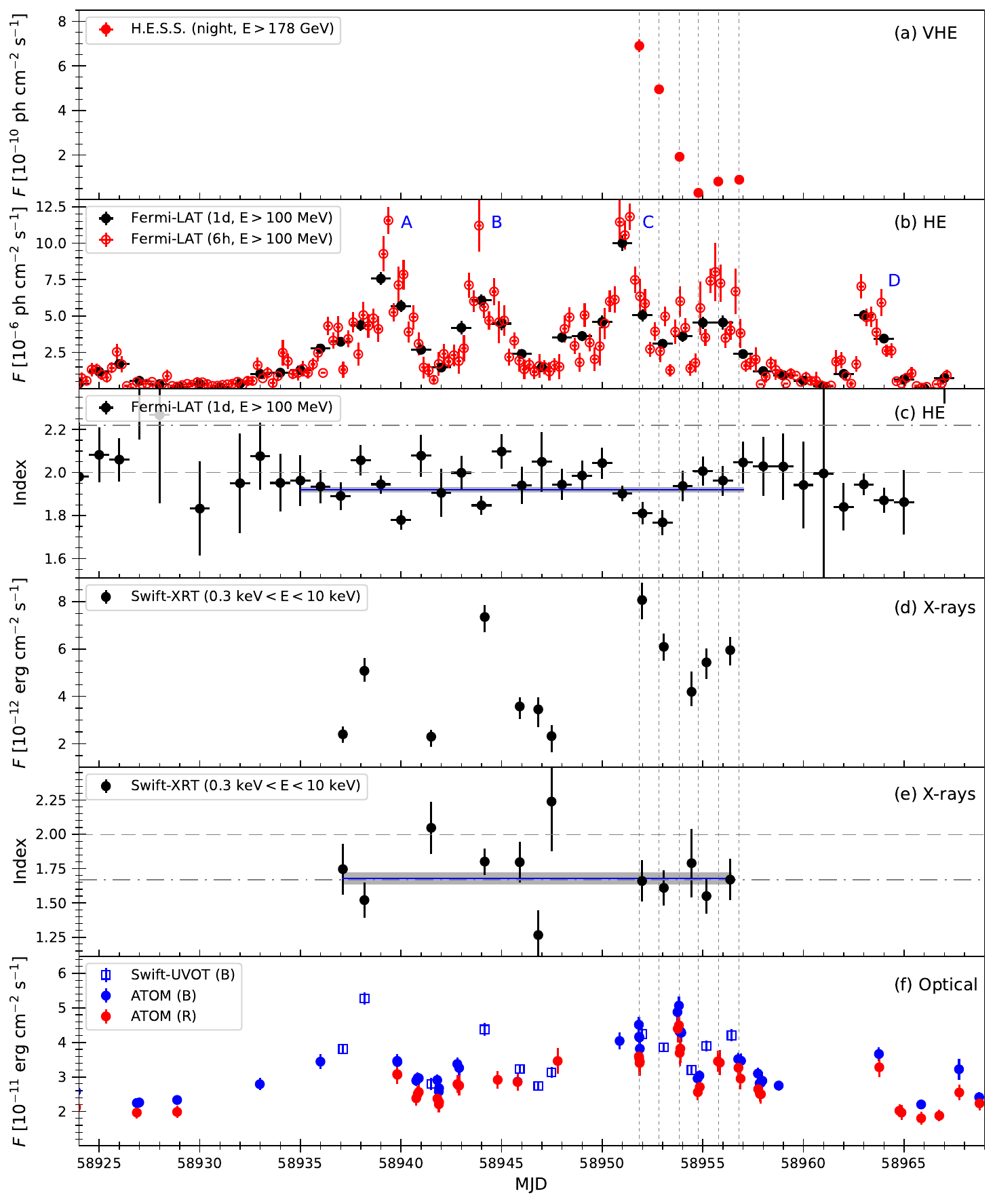}
\caption{MWL light curves in the VHE \g-ray band from \hess\ (a), in the HE \g-ray band from \fermi\ (b), in the X-ray band from \textit{Swift}-XRT (d), and in the optical bands from \textit{Swift}-UVOT and ATOM (f). Panels (c) and (e) show the spectral index in the HE \g-ray and X-ray bands, respectively. The blue letters in panel (b) label the four separate peaks in the HE \g-ray light curve. Horizontal gray dashed lines in panels (c) and (e) mark an index value of 2.0, horizontal gray dash-dotted lines mark the long-term averages, while the blue horizontal lines and the gray regions mark the average during the flare and its error. The vertical dotted lines mark the H.E.S.S. observation times. The data points are individual observations unless otherwise noted.
}
\label{fig:lc_mwl_full}
\end{figure*}
\begin{figure*}
\centering
\includegraphics[width=0.98\textwidth]{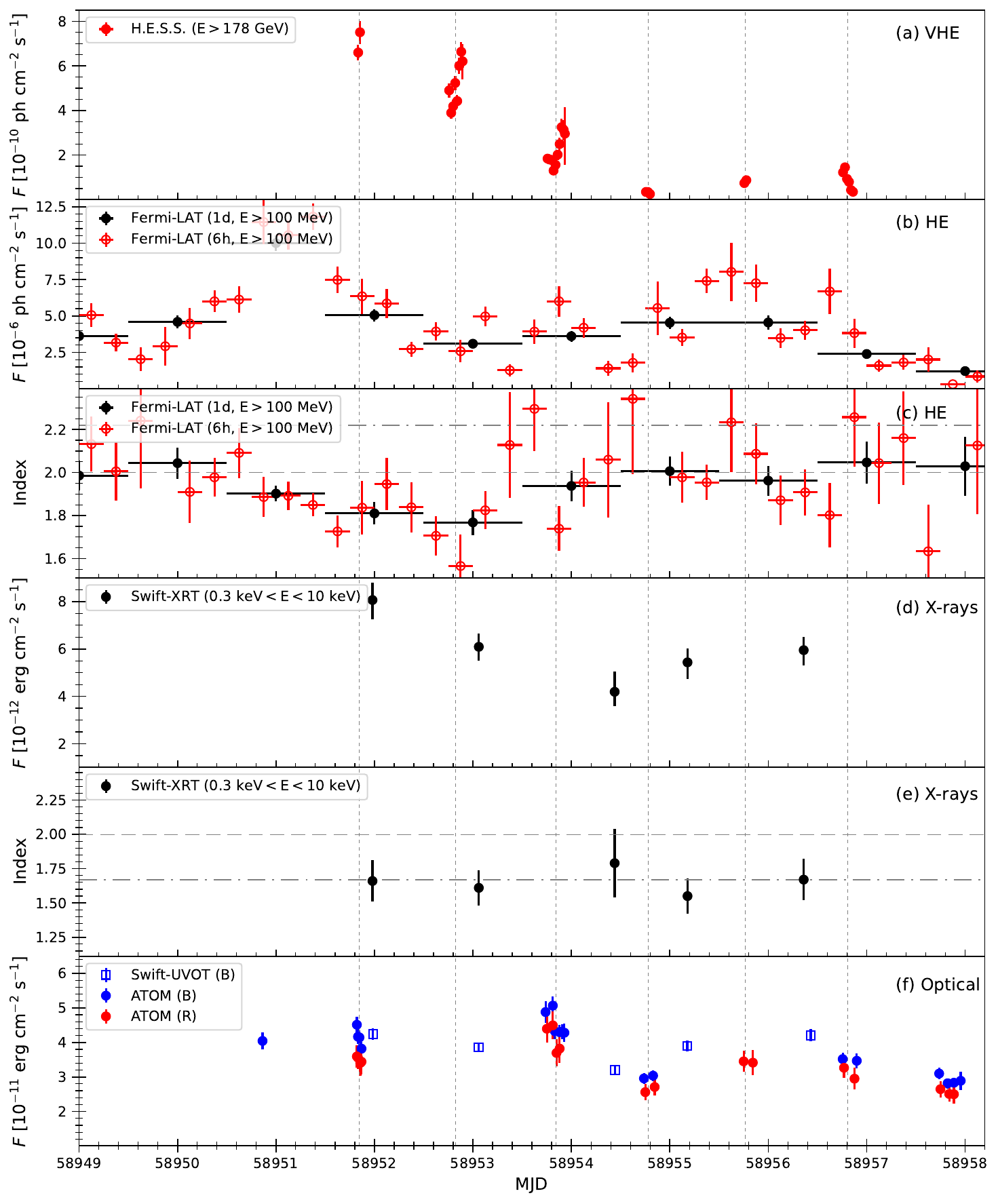}
\caption{Same as Fig.~\ref{fig:lc_mwl_full}, but zoomed in on the nights with \hess\ observations. 
}
\label{fig:lc_mwl_zoom}
\end{figure*}
%

%%%%%%%%%%%%%%%%%%%%%%%%%%%%%%%%%%%%%%%%%%%%%%%%%%
\subsection{VHE \g\ rays} \label{sec:vhe}
\begin{table}
\caption{Observation times with H.E.S.S. in the VHE \g-ray band.}
\label{tab:vheobs}
\begin{tabular}{llll}
\hline\hline
Data Set & MJD start & MJD stop & $T_{\rm obs}$ [h] \\
\hline
Full    & 58951.8250 & 58956.8706 & 13.1 \\
Night 1 & 58951.8250 & 58951.8604 & 0.7 \\
Night 2 & 58952.7493 & 58952.8985 & 3.3 \\
Night 3 & 58953.7487 & 58953.9366 & 4.1 \\
Night 4 & 58954.7482 & 58954.8082 & 1.4 \\
Night 5 & 58955.7476 & 58955.7873 & 0.9 \\
Night 6 & 58956.7496 & 58956.8706 & 2.7 \\
\hline
\end{tabular}
\end{table}
\hess\ is an Imaging Air Cherenkov Telescope array located in the Khomas Highland in Namibia ($\mathrm{23^{\circ}16'18''}$S, $\mathrm{16^{\circ}30'00''}$E), at an altitude of 1800 m above sea level.
\hess\ consists of four telescopes (CT1-4) with a mirror-dish diameter of 12\,m, and one larger telescope (CT5) with a 28\,m-mirror dish. The cameras of CT1-4 were refurbished in 2016, while an advanced FlashCam prototype \citep{werner+17} has been routinely operating in the CT5 telescope since the end of 2019.

\hess\ observations of \source\ were triggered following alerts on an active state at high energies reported by the \texttt{FLaapLUC} pipeline \citep{2018A+C....22....9L} on March 28, 2020. 
Unfavorable weather conditions prevented \hess\ observations before April 12, 2020. In the following 6 nights until April 17, 2020, \source\ was observed with \hess\ for a total of 13.1 hours (see Tab.~\ref{tab:vheobs} for details). All data passed the standard quality selection \citep{hess06}. 
The observations were performed in wobble mode with \source\ positioned at an offset of 0.5$^{\circ}$ from the center of the camera.

\hess\ data were analyzed using 
events that were received by the larger telescope, CT5, in monoscopic mode enabling to probe events at the lowest possible energies in the VHE range. The events were reconstructed using the neural-network-based chain described in \cite{murach+15} with an image amplitude cut of 250 p.e. \citep[see also][]{puelhofer+22}, 
which achieves a smaller energy bias and a better \g/hadron separation, making the results more reliable. 
The cross-checks provide consistent results and are presented in \ref{app:xcheck}.

The 
analysis, conducted with the ring background method \citep{berge+07}, yields a total significance of $103\sigma$ following \citet[Eq.~(17)]{LiMa1983}, with 8,067 excess \g-like events, 11,237 events in the source region, 35,120 events in the background region, a solid angle ratio between the source and the background region of $11.1$, and a signal-to-noise ratio of $2.5$. The reflected background method \citep{berge+07} was used for the spectral analysis. The energy threshold of the spectrum is $\sim$$160\,$GeV.
The total and night-wise spectra were 
fitted with a power-law 

\begin{equation}
    \frac{\td{N}}{\td{E}} = N_0 \left(\frac{E}{E_0}\right)^{-\Gamma} \label{eq:pwl}
\end{equation}
and a log-parabola 

\begin{equation}
    \frac{\td{N}}{\td{E}} = N_0 \left(\frac{E}{E_0}\right)^{-\Gamma - \beta \ln{(E/E_0)}} \label{eq:log-parabola}
\end{equation} 
where $N_0$ is the differential photon flux normalization at reference energy $E_0=300\,$GeV, $\Gamma$ is the photon index, and $\beta$ is the curvature index. The spectrum is considered to be curved if the log-likelihood ratio test\footnote{This is defined as $\Delta \text{TS}=-2\ln{\mathcal{L}_{\rm lp}/\mathcal{L}_{\rm pl}}$, where $\Delta \text{TS}$ is the test statistics and $\mathcal{L}_{\rm lp/pl}$ are the likelihoods of a log-parabola and power-law fit, respectively \citep{mattox+96}.} between the log-parabola and the power-law is $\Delta \text{TS}\geq 9$ corresponding to about $3\sigma$ \citep{wilks38}. This is the case for the total spectrum, as well as Nights 1, 2, and 3.
The spectral results, and the $\Delta \text{TS}$ are listed in Tab.~\ref{tab:vhemono}.

The overall energy spectrum is shown in Fig.~\ref{fig:spec_hess_full}, while the energy spectra for individual nights are presented in Fig.~\ref{fig:spec_hess_night}. In each case, forward-folded spectral points (rebinned to at least $2\sigma$ per point) 
with statistical error bars, as well as butterflies with statistical (dark shades) and systematic (light shades; see below) errors are plotted. Also shown are the spectral points  
corrected for absorption in the extragalactic background light (EBL) using the model of \cite{Franceschini+08}. 
Performing the log-likelohood ratio test on the spectra including the EBL correction, reveals that only the total spectrum is mildly curved ($\Delta \text{TS} = 9.6$), while the night-wise spectra are not ($\Delta \text{TS}\leq 4$). Hence, most of the observed curvature is due to the EBL absorption, but the intrinsic VHE \g-ray spectra are soft (i.e., $\Gamma>2$).

Light curves have been calculated  
with the energy threshold set at an energy of $178\,$GeV, which is the highest energy threshold of the night-wise spectra (see Tab.~\ref{tab:vhemono}). Light curves are shown in Fig.~\ref{fig:lc_mwl_full}(a) for night-wise bins and in Fig.~\ref{fig:lc_mwl_zoom}(a) for individual observation times. The average flux is $1.5\E{-10}\,$ph\,cm$^{-2}$s$^{-1}$, which corresponds to $60\%$ of the Crab Nebula flux above the same threshold \citep{hess06}.

In order to derive the systematic errors, the choice of analysis cuts, the acceptance description of CT5, as well as the atmospheric transparency are considered following \cite{hess2022Sci}. The dominating systematic uncertainty, however, stems from the unknown absolute energy scale. This was evaluated using the open-source software \texttt{gammapy} \citep[version 1.2,][]{2023A+A...678A.157D,acero_2024_10726484}, and was done by globally rescaling the energy attributed to the data sets with a Monte Carlo simulation drawing from a Gaussian distribution centered on unity 
with a width corresponding to the assumed accuracy of the energy scale of 10\%. With these rescaled data sets, the spectra were rederived. This procedure was repeated 10,000 times, and the resulting uncertainties on the spectral parameters are taken as the systematic errors from this effect. The final systematic errors, as given in Tab.~\ref{tab:vhemono}, 
have been derived by summing the various contributions in quadrature.

\begin{table*}
\caption{VHE \g-ray observed spectral data for spectra shown in Figs.~\ref{fig:spec_hess_full} and~\ref{fig:spec_hess_night}. $E_{\rm thr}$ is the energy threshold of the data set. The last column gives the preference of the log-parabola spectrum over the power-law spectrum, $\Delta \text{TS}$.
}
\label{tab:vhemono}
\begin{tabular}{llllll}
\hline\hline
Data Set & $N_0$ & $\Gamma$ & $\beta$ & $E_{\rm thr}$ &  $\Delta \text{TS}$ \\
 & [$10^{-10}\,$ph\,cm$^{-2}$s$^{-1}$TeV$^{-1}$] &  &  & [GeV] & \\
\hline
Full    & $6.7\pm 0.2\stat\pm 3.1\sys$   & $3.89\pm 0.06\stat\pm 0.3\sys$ & $1.1\pm 0.2\stat\pm 0.1\sys$ & $162$ & $81.4$ \\
Night 1 & $19.1\pm 2.0\stat\pm 10.3\sys$ & $4.4\pm 0.2\stat\pm 0.5\sys$   & $1.9\pm 0.8\stat\pm 0.1\sys$ & $178$ & $17.4$ \\
Night 2 & $12.9\pm 0.6\stat\pm 6.1\sys$  & $3.88\pm 0.08\stat\pm 0.3\sys$ & $0.9\pm 0.2\stat\pm 0.1\sys$ & $162$ & $36.8$ \\
Night 3 & $4.9\pm 0.4\stat\pm 2.4\sys$   & $3.9\pm 0.1\stat\pm 0.3\sys$   & $0.7\pm 0.3\stat\pm 0.1\sys$ & $162$ & $10.3$ \\
Night 4 & $0.7\pm 0.2\stat\pm 0.3\sys$   & $3.5\pm 0.5\stat\pm 0.4\sys$   & --                           & $162$ & $1.1$ \\
Night 5 & $1.4\pm 0.3\stat\pm 0.9\sys$   & $4.6\pm 0.4\stat\pm 0.4\sys$   & --                           & $162$ & $1.3$ \\
Night 6 & $1.8\pm 0.2\stat\pm 1.0\sys$   & $4.2\pm 0.2\stat\pm 0.3\sys$   & --                           & $162$ & $0.5$ \\
\hline
\end{tabular}
\end{table*}
%

%%%%%%%%%%%%%%%%%%%%%%%%%%%%%%%%%%%%%%%%%%%%%%%%%%
\subsection{HE \g\ rays} \label{sec:he}
%
% JP
\fermi\ is a pair conversion instrument designed to detect \g\ rays with energies from 20\,MeV to above 300\,GeV \citep{dd09}.
Public data from \fermi\ were analysed in time intervals contemporaneous to the \hess\ datasets. Two different types of data selection have been carried out. For a global picture of the flare, an analysis was performed on \fermi\ data encompassing the \hess\ observation campaign described in Section~\ref{sec:vhe}.
The \fermi\ P8R3 event selection was used \citep{bruel+18}, with the \texttt{SOURCE} event class (\textit{evclass=128}) and joint FRONT$+$BACK event types (\textit{evtype=3}). The analysis was performed with the \texttt{Fermitools} (version 2.2.0) software\footnote{\url{https://github.com/fermi-lat/Fermitools-conda/wiki}}, which are the official \textit{Science Tools} provided by the Fermi Science Support Center\footnote{\url{https://fermi.gsfc.nasa.gov/ssc/}}. A binned analysis was employed between energies of 100\,MeV and 500\,GeV for a region of interest (ROI) with radius 10$^{\circ}$ centred at the nominal position of \source. To reduce contamination from the Earth limb in \g\ rays, a zenith angle cut of 90$^{\circ}$ was applied. All sources belonging to the 4FGL-DR3 falling within 15$^{\circ}$ around \source\ have been accounted for in the likelihood analysis.

The iterative likelihood fitting procedure adopted here is detailed in \citet[][Section~3.1]{2018A+C....22....9L}, and is summarized as follows. In a first step, for each source included in the ROI model, their respective spectral parameters are taken from the 4FGL-DR3 catalogue as seed inputs. Their parameters are frozen if no hint of detection is found with a test statistics\footnote{The TS is defined here as twice the difference of log-likelihoods of the optimised ROI model with and without the source included, TS=$-2 (\ln{\mathcal{L}_1}-\ln{\mathcal{L}_0})$.} of $\text{TS}<9$ and if the predicted number of photons from that source contributes to less than 5\% of the total photon counts within the ROI. In a second step, only the spectral parameters of sources lying within 3$^{\circ}$ from \source\ are left free to vary. In a last iteration, only the spectral parameters of \source\ are left free to vary, all other source parameters are fixed. The normalization of the Galactic and isotropic background templates\footnote{using the diffuse models \texttt{gll\_iem\_v07} and \texttt{iso\_P8R3\_SOURCE\_V3\_v1} respectively, see \href{https://fermi.gsfc.nasa.gov/ssc/data/access/lat/BackgroundModels.html}{https://fermi.gsfc.nasa.gov/ssc/data/access/lat/BackgroundModels.html}.} are left as additional free parameters.

Figures~\ref{fig:lc_mwl_full}(b) and~\ref{fig:lc_mwl_zoom}(b) present the light curve at HE \g\ rays of \source\ modelled using a simple power-law spectrum, %:
Eq.~\eqref{eq:pwl}, with the reference energy, $E_0=903$\,MeV, fixed at the value from the 4FGL-DR3 catalogue.
Two different time binnings are used: 1\,d and 6\,h. The latter better illustrates the fast variability from \source\ at the expense of sparser photon statistics per point. The photon index is left free to vary to probe for any spectral variability during this active state and is shown in panel (c) of Figs.~\ref{fig:lc_mwl_full} and~\ref{fig:lc_mwl_zoom}.

Individual \fermi\ spectra for each \hess\ observation night have been derived and are shown in Fig.~\ref{fig:spec_model}. The 12-hour data integration employed, centered on the \hess\ observations, is a trade-off between integrating long enough to accumulate enough photon statistics to decrease statistical errors on the fitted spectral parameters while keeping it short enough to still represent the same state of activity as the \hess\ one. For each night, two spectral hypotheses are tested, namely a power-law (see Eq.~\ref{eq:pwl}) or a log-parabola spectrum (see Eq.~\ref{eq:log-parabola}) %:
with the preferred spectral shape chosen as in Sec.~\ref{sec:vhe}.
The analysis results for each night are summarized in Tab.~\ref{tab:fermi}. Even though not studied in this work, \fermi\ results are also prone to systematic effects. Specifically concerning the absolute energy scale, it has been estimated to an accuracy of ${}_{-5}^{+2}\%$ \citep{2012APh....35..346A}.

\begin{table*}
\begin{threeparttable}
\caption{\fermi\ spectral analysis results for each \hess\ night.}\label{tab:fermi}
\begin{tabular}{llllll}
\hline\hline
Central time & Detection significance & $F_{0.1-500\,\mathrm{GeV}}$ & $\Gamma$ & $\beta$ & $E_\mathrm{max}$$^{\text{a}}$ \\
 $[$MJD$]$ & [$\sigma$] & [$10^{-6}$\,ph\,cm$^{-2}$\,s$^{-1}$] &  & & [GeV]\\ %UTC
\hline
58951.8333 & 28.2 & $6.29 \pm 0.68$ & $1.666 \pm 0.082$ & $0.126 \pm 0.050$ & 20.38\\ %2020-04-12 20:00
58952.8333 & 23.0 & $3.87 \pm 0.57$ & $1.526 \pm 0.110$ & $0.089 \pm 0.049$ & 36.96\\ %2020-04-13 20:00
58953.8333 & 19.7 & $4.82 \pm 0.65$ & $1.859 \pm 0.101$ & --                & 20.33\\ %2020-04-14 20:00
58954.7917 & 10.4 & $3.43 \pm 0.73$ & $2.284 \pm 0.246$ & --                &  4.63\\ %2020-04-15 19:00
58955.75   & 15.6 & $7.41 \pm 1.10$ & $2.229 \pm 0.175$ & --                &  4.36\\ %2020-04-16 18:00
58956.7917 & 14.5 & $3.68 \pm 0.65$ & $1.994 \pm 0.136$ & --                &  9.93\\ %2020-04-17 19:00
\hline
\end{tabular}
\begin{tablenotes}
\footnotesize
\item $^{\text{a}}$~Energy of the highest energy photon associated with \source\ with a probability higher than 95\%.
\end{tablenotes}
\end{threeparttable}
\end{table*}

%%%%%%%%%%%%%%%%%%%%%%%%%%%%%%%%%%%%%%%%%%%%%%%%%%
\subsection{X rays} \label{sec:xrays}
%
% AW
In 2020 the X-Ray Telescope (XRT) onboard the Neil Gehrels Swift Observatory \citep[for details see][]{Swift} observed PKS\,0903-57 12 times (ObsID of 00033856009-00033856020). 
All observations were performed in the photon counting mode in the energy range of 0.3–10\,keV. 
The data analysis of the X-ray observations was performed using the HEASOFT software (version 6.32 with CALDB version 20211108). 
All data were binned to have at least  30  counts per bin. 
For each of the observations, the analysis considered all photons within a circular region with a radius of 5$^{\prime\prime}$. The background was estimated from a separate circular region positioned near the source but in an area free from contamination by signals from nearby sources.
Spectral results for the observations during the \hess\ observation window are reported in Tab.~\ref{tab:xrt} and shown in Fig.~\ref{fig:spec_model}.
Energy fluxes have been derived by fitting each single observation with a power-law model with a  Galactic absorption value of {\it N}$_H$= 2.6$\times$10$^{21}$\,cm$^{-2}$ \citep[HI4PI Map,][]{HI4PI}. 
The X-ray light curve is shown in Figs.~\ref{fig:lc_mwl_full}(d) and~\ref{fig:lc_mwl_zoom}(d), and the corresponding photon index in Figs.~\ref{fig:lc_mwl_full}(e) and~\ref{fig:lc_mwl_zoom}(e), respectively.

\begin{table*}
\caption{Summary of the \textit{Swift}-XRT spectral analysis results for single observations during the \hess\ observation window.
}
\label{tab:xrt}
\begin{tabular}{llllll}
\hline\hline
ID of observation & Time of observation & Duration  &  N    & $\Gamma$ & $\chi^2(n_{d.o.f.})$   \\
  & [MJD]  &  [s] & [$10^{-4}$\,cm$^{-2}$\,s$^{-1}$\,keV$^{-1}$]   &    &\\ %UTC
\hline
00033856016 & 58951.9789 & 1990  &  $15.0 \pm 1.6$ & $1.66 \pm 0.15$ & 16.2(17) \\ %2020-04-12 23:29:41
00033856017 & 58953.0511 & 1991  &  $15.9 \pm 1.6$ & $1.61 \pm 0.13$ & 22.2(20) \\ %2020-04-14 01:13:35
00033856018 & 58954.4365 &  777  &  $14.4 \pm 2.6$ & $1.79 \pm 0.25$ & 23.7(25) \\ %2020-04-15 10:28:35
00033856019 & 58955.1712 & 2024  &  $16.0 \pm 1.6$ & $1.55 \pm 0.13$ & 17.2(19) \\ %2020-04-16 04:06:34
00033856020 & 58956.3601 & 1983  &  $13.4 \pm 1.5$ & $1.67 \pm 0.15$ & 14.3(16) \\ %2020-04-17 08:38:30
\hline
\end{tabular}
\end{table*} 

%%%%%%%%%%%%%%%%%%%%%%%%%%%%%%%%%%%%%%%%%%%%%%%%%%
\subsection{Optical data} \label{sec:uvot}
%
% AW, MZ
%
%UVOT
The  UltraViolet and Optical Telescope (UVOT) observations were performed simultaneously with the X-ray ones aboard \textit{Swift}.  
For the analysis of each of these observations all photons in a circular region with a radius of 5$^{\prime\prime}$ were taken into account.
The background was determined also within a circular region located near the source region.
The instrumental magnitudes and the corresponding fluxes were calculated with the \verb|uvotsource| procedure, while
the conversion into flux units was done using the factors taken from \cite{Poole08}.
All measured fluxes were dereddened using $E(B-V) =  0.2827$\,mag \citep{Schlafly11}  and  $A_{\lambda}/E(B-V)$, as provided by \cite{Giommi06}.
The B-band light curve is shown in Figs.~\ref{fig:lc_mwl_full}(f) and \ref{fig:lc_mwl_zoom}(f).

%ATOM
The Automatic Telescope for Optical Monitoring (ATOM) is an optical telescope with 75\,cm aperture located on the \hess\ site \citep{hauser+04}. ATOM observed \source\ throughout the flaring period employing BR filters. During the flare the cadence was increased to obtain more than one observation per filter per night.
The ATOM data were analysed using the fully automated ATOM Data Reduction and Analysis Software and their quality has been checked manually. The resulting flux was calculated via differential photometry using five custom-calibrated secondary standard stars in the same field of view. Extinction correction was done as for \textit{Swift}-UVOT. The light curves are shown in Figs.~\ref{fig:lc_mwl_full}(f) and \ref{fig:lc_mwl_zoom}(f).

%%%%%%%%%%%%%%%%%%%%%%%%%%%%%%%%%%%%%%%%%%%%%%%%%%
\subsubsection{The star} \label{sec:star}
Instruments like \textit{Swift}-UVOT and ATOM cannot separate \source\ and \thestar. The latter's influence is clearly visible in Fig.~\ref{fig:spec_model} through the narrow bump in the data at NIR to UV energies. 

In order to assess the influence of \thestar\ on the spectrum, the Gaia DR3 catalog \citep{gaia16,gaia23} was consulted. It lists \source\ with catalog entry \texttt{Gaia DR3 5303862539845306624}, while \thestar\ has the entry \texttt{Gaia DR3 5303862535547465856}. The Gaia catalog provides \thestar's $G_{RP}$ and $G_{BP}$ magnitudes as $G_{RP}=15.0$ and $G_{BP}=16.0$, which have been converted to energy fluxes using the AB conversion. The resulting flux points are shown in Fig.~\ref{fig:spec_model} as blue filled asterisks. They align well with the archival data (gray points). Additionally, the Gaia DR3 magnitudes were corrected for Galactic extinction using the same method as for the \textit{Swift}-UVOT and ATOM data. The resulting extinction-corrected flux points are shown as red asterisks in  Fig.~\ref{fig:spec_model}. These indicate similar spectral trends as the data from \textit{Swift}-UVOT and ATOM taken during the flare. This suggests that the optical data of the flare is indeed strongly influenced by \thestar's emission.

The extinction correction following \cite{Schlafly11} strictly holds only for sources located beyond the Milky Way, which is not true for \thestar. The optical fluxes from the flare are thus somewhat overcorrected, although the magnitude of this overcorrection is difficult to quantify. As it is not possible at this point to remove \thestar's influence from the data, this is considered a reasonable compromise in order to obtain a limit on the blazar's synchrotron flux during the flare (see Sec.~\ref{sec:modelstrat}).

The Gaia DR3 catalog gives a $G$ magnitude for \source\ of $G=17.9$. This is converted to energy fluxes using the same steps as above. The result is shown in Fig.~\ref{fig:spec_model} as open asterisks with the same color coding as for \thestar. The fluxes align well with the archival data in the (far-)IR bands suggesting that they are representative of the blazar's low state.

%
%%%%%%%%%%%%%%%%%%%%%%%%%%%%%%%%%%%%%%%%%%%%%%%%%%%%%%%%%%%%%%%%%%%%%%%%%%%%%%%%%%%%%%%%%
\section{Flux and spectral variability} \label{sec:varia}

\begin{table*}
\caption{Fractional variabilitly, $F_{\rm var}$, for different energy bands and the full flaring period (MJD 58932--58966) and focusing on the \hess\ observation window (MJD 58951--58957). The ``B band'' is the combination of the \textit{Swift}-UVOT/B and ATOM/B light curves.
}
\label{tab:Fvar}
\begin{tabular}{lcc}
\hline\hline
Observatory & MJD 58932--58966 & MJD 58951--58957 \\
\hline
\hess\ (night) & -- & $1.02\pm0.02$ \\ 
\hess\  & -- & $0.79\pm0.02$ \\ 
\fermi\ (1d) & $0.69\pm0.02$ & $0.45\pm0.03$ \\ 
\fermi\ (6h) & $0.79\pm0.02$ & $0.43\pm0.05$ \\ 
\textit{Swift}-XRT & $0.40\pm0.03$ & $0.21\pm0.05$ \\ 
\textit{Swift}-UVOT/B & $0.19\pm0.01$ & $0.10\pm0.02$ \\ 
ATOM/B & $0.20\pm0.01$ & $0.14\pm0.01$ \\ 
B band & $0.24\pm0.007$ & $0.13\pm0.01$ \\
ATOM/R & $0.19\pm0.02$ & $0.11\pm0.02$ \\ 
\hline
\end{tabular}
\end{table*}

The VHE \g-ray light curve is shown in Fig.~\ref{fig:lc_mwl_full}(a) for night-wise binning and in Fig.~\ref{fig:lc_mwl_zoom}(a) for individual observations. The night-wise binning shows an overall decreasing flux in the first four nights. Interestingly, the intra-night binning is indicative of rising trends in Nights 2 and 3, while Night 6 shows a decreasing trend. The variability time scale in these nights is on the order of $\sim 5\,$h. This is estimated by dividing the average flux of the individual night by the slope of a linear fit to the night's flux points. While the spectral parameters in Tab.~\ref{tab:vhemono} 
may be suggestive of spectral variability, the large systematic errors prevent any such conclusion.

The overall HE \g-ray light curve in Fig.~\ref{fig:lc_mwl_full}(b) shows four distinct flares (labeled A, B, C, D) with the peak C being surrounded by elevated plateau phases. Both plateaus display variable fluxes, but no distinct peak like the main ones -- even though \cite{Shah+21} interpret the flux evolution around Night 5 as an additional peak. Measuring the variability time scale between subsequent flux points, $F_i(t_i)$, using

\begin{align}
    t_{\rm var} = \frac{F_1+F_2}{2} \frac{t_2-t_1}{|F_2-F_1|}
    \label{eq:tvar},
\end{align}
the minimum variability time scale, $t_{\rm var,min}$, in the HE band is constrained to $\sim 6\,$h. While this depends on the choice of binning, it is in line with the result of \cite{Shah+21}, who conducted a more thorough investigation of the HE light curve. The minimum time scale is comparable to the one found in the VHE band.

The evolution of the HE spectral index is shown in Fig.~\ref{fig:lc_mwl_full}(c) indicating a spectrum that is consistently harder than the 4FGL-DR3 spectrum (marked by the solid gray horizontal line). During flares A, B and C, the average index is $1.92\pm0.01$ (blue horizontal line), but a constant index is ruled out with $3.2\sigma$. Stretching the considered time frame to also include flare D, keeps the average index unchanged, but makes it compatible with a constant (rejection significance $2.5\sigma$).
Moreover, while during flare B the lowest spectral index value coincides with the flux peak, the spectral hardening is delayed with respect to the flux peak in flares A and C. 
There is no obvious spectral evolution during flare D.

The comparison of the VHE fluxes with the HE flux and index on the falling side of flare C suggests an interesting evolution of the \g-ray component. While the nightwise VHE flux shows a decreasing trend in Nights 1 to 4, the corresponding 6\,h HE fluxes are almost equal except for a low flux in Night 2. However, the HE index is $<2$ in Night 1 to 3, but $>2$ in Night 4. 
In Night 5, the VHE flux is only mildly increased compared to Night 4, as the almost equally high HE flux (compared to previous nights) is also accompanied by a soft spectrum. Night 6 shows a decreasing trend in both the VHE and HE fluxes, again with a soft HE spectrum.
As the VHE spectrum even after the correction for the EBL is consistently soft ($\Gamma>2$), the described behavior indicates not just a spectral break between the HE and VHE band, but also an evolution in the break energy. This will be discussed further below.

The X-ray flux [Figs.~\ref{fig:lc_mwl_full}(d) and \ref{fig:lc_mwl_zoom}(d)] seems to follow mostly the HE flux evolution, even though there is a delay during flare C seemingly matching the VHE flux evolution. The evolution of the X-ray spectral index, shown in panel (e) of both figures, is compatible with a constant during the flaring period with an average value of $1.68\pm 0.04$. This is in line with the historical value. This behavior resembles that of other FSRQs \citep[e.g.,][]{magic17,hess19,hess23}.

The optical flux [panel (f) in Figs.~\ref{fig:lc_mwl_full} and \ref{fig:lc_mwl_zoom}] seems to correlate with the HE flux in flares A, B and D, but not so much during and after flare C. While the gap of observations in the run-up to flare C may play a role (even though the first point after the gap is during the HE peak flux), this still suggests that the synchrotron flux was less affected than the X-ray to \g-ray flux.

The flux variability in an energy band can also be characterized with the so-called fractional variability, $F_{\rm var}$, defined as \citep{vaughan+03}

\begin{align}
    F_{\rm var} = \sqrt{\frac{S^2-\overline{\sigma_{\rm err}^2}}{\overline{F}^2}}
    \label{eq:fvar}.
\end{align}
The associated error is \citep{poutanen+08}

\begin{align}
    \Delta F_{\rm var} = \sqrt{F_{\rm var}^2 + {\rm err}(\sigma_{NXS}^2)} - F_{\rm var}
    \label{eq:dfvar}
\end{align}
with the error on the normalized excess variance

\begin{align}
    {\rm err}(\sigma_{NXS}^2) = \sqrt{\frac{2 \overline{\sigma_{\rm err}^2}}{N\overline{F}^4} \left( 2F_{\rm var} \overline{F}^2 + \overline{\sigma_{\rm err}^2} \right)}
    \label{eq:errnormexcvar}.
\end{align}
In Eqs.~\eqref{eq:fvar} to~\eqref{eq:errnormexcvar}, $S^2$ and $\overline{F}$ are the variance and arithmetic mean of the light curve, and $\overline{\sigma_{\rm err}^2}$ is the mean square of the measurement errors. The fractional variability for each energy band is given in Tab.~\ref{tab:Fvar} for the entire flaring period as well as limited to the \hess\ observation range. While the incompleteness of most energy bands makes a comparison tricky, there is a general trend that $F_{\rm var}$ increases with energy. The low $F_{\rm var}$ in the optical domain is probably a consequence of \thestar. \cite{Shah+21} derived the $F_{\rm var}$ of all UVOT bands during the flare showing an increasing trend with energy in these bands, which can be related to the waning influence of \thestar\ towards the UV domain. Combining the UVOT and ATOM B-band light curves provides a slightly higher $F_{\rm var}$ than considering the light curves separately suggesting that the combination of the two data sets provides a more complete sample of the variability.

During the \hess\ observation window, the $F_{\rm var}$ is reduced compared to the full window. This is expected as a shorter observation window produces a smaller $F_{\rm var}$. Additionally, the highest and lowest fluxes from the full flaring period are missing. Nevertheless, the trend of higher $F_{\rm var}$ for higher energies persists, as the VHE \g-ray light curve exhibits the highest fractional variability of all bands during that time. It is also interesting to note the influence of the time-binning on the fractional variability. While the HE \g-ray light curve during the full time range exhibits a higher $F_{\rm var}$ for shorter time binning, there is no such trend during the \hess\ window. Meanwhile, the VHE \g-ray light curve exhibits a higher fractional variability for the longer time integration. 
This is in line with the elaborate study of \cite{schleicher+19} pointing at the influence of the incompleteness of the data sets on the fractional variability.

Figure~\ref{fig:spec_model} shows the MWL SEDs collected near-simultaneously to the \hess\ observations. 
The data recorded with the instruments aboard \textit{Swift} are only strictly simultaneous with the \hess\ and ATOM observations for Nights 1 and 2, while they are about 12 hours behind for Nights 3 and 4 and less than 12 hours ahead for Night 6; see Fig.~\ref{fig:lc_mwl_zoom}. According to Tab.~\ref{tab:Fvar}, the variability in the X-ray and optical range is low during the \hess\ observation window. Thus, the \textit{Swift} spectra can be used in the mentioned nights without strongly influencing the interpretation.

While the optical flux and spectrum barely change owing in part to \thestar, the \g-ray spectrum changes significantly. In comparison to the 4FGL spectrum, the flux increases by more than an order of magnitude and the peak energy 
shifts significantly upward. From the 4FGL spectrum, the average peak position is at roughly $240\,$MeV. Fitting the night-wise observed HE and VHE spectra with a log-parabola model, one finds that the peak is attained at $(6.4\pm1.6)\,$GeV and $6.6\pm1.5\,$GeV during Nights 1 and 2, respectively; more than an order of magnitude higher. Nights 3 and 6 also exhibit higher peak energies compared to the 4FGL spectrum [$(3.8\pm1.2)\,$GeV and $(3.2\pm1.8)\,$GeV, respectively]. Nights 4 and 5 [$(0.2\pm0.2)\,$GeV and $(1.6\pm1.3)\,$GeV, respectively] are not well constrained and compatible within errors with the long-term value. The evolution in peak energy is in line with the spectral changes described above.

%
%%%%%%%%%%%%%%%%%%%%%%%%%%%%%%%%%%%%%%%%%%%%%%%%%%%%%%%%%%%%%%%%%%%%%%%%%%%%%%%%%%%%%%%%%
\section{Leptonic modeling} \label{sec:model}
Using the observations described in the previous sections, a leptonic modeling is conducted of the data taken during the nights of the \hess\ observations. Firstly, the constraints set by the data are presented, followed by the modeling strategy. Lastly, the results are discussed.
In the following, primed quantities are considered in the frame of the host galaxy, unprimed quantities in the comoving frame of the jet, and quantities marked with a subscript \textit{obs} are taken in the observer's frame

%%%%%%%%%%%%%%%%%%%%%%%%%%%%%%%%%%%%%%%%%%%%%%%%%%
\subsection{Constraints} \label{sec:modelconst}

The observed variability time scale of $\sim 5\,$h suggests a radius of the emission region of

\begin{align}
    R< 2.1\E{16}\,\mbox{cm}\, \est{t_{\rm var,obs}}{5\,\mbox{h}}{} \est{\delta}{50}{}
    \label{eq:Rlim}
\end{align}
with the Doppler factor, $\delta$.
Assuming a constant opening angle of the conical jet, $\Theta/\Gamma_b$ with $\Theta\sim 0.2$ \citep{Jorstad+17} and the bulk Lorentz factor, $\Gamma_b$, the limit on $R$ implies a limit on the distance from the black hole, $d$:

\begin{align}
    d &= \frac{R}{\tan{(\Theta/\Gamma_b)}} \approx \frac{\Gamma_b R}{\Theta} \nonumber \\
    &< 5.3\E{18}\,\mbox{cm}\, \est{t_{\rm var,obs}}{5\,\mbox{h}}{} \est{\delta\,\Gamma_b}{2500}{} \est{\Theta}{0.2}{-1}.
    \label{eq:dlim}
\end{align}
This limit assumes that the emission region fills the width of the jet. This is not necessarily true as shown in other flares \citep[e.g.,][]{hess+21}.

The Doppler factor can be constrained by demanding that the emission region is optically thin to the observed TeV radiation. This argument only considers jet-intrinsic photon fields. Optical depth of \g\ rays with respect to external photon fields will be discussed below. The detailed steps to derive the limit on $\delta$ are provided in \ref{app:doppler} resulting in a minimum value of 

\begin{align}
    \delta > 8.9 \est{\Ego}{1\,\mbox{TeV}}{1/5} \est{R}{10^{16}\,\mbox{cm}}{-1/5} \est{(\nF)_{\rm syn}}{2.5\E{-12}\,\frac{\mbox{erg}}{\mbox{cm}^2\,\mbox{s}}}{1/5}
    \label{eq:tauopt},
\end{align}
where $\Ego$ is the observed \g-ray energy, and $(\nF)_{\rm syn}$ is the low-state synchrotron energy flux. If the synchrotron flux increased by an order of magnitude during the flare, the limit would rise to $\delta\gtrsim 15$.

A consequence of the lower flux variations in the X-ray band compared to the HE \g-ray band as well as the stability of the X-ray spectral index, is that the X-ray spectrum does not connect well to the HE \g-ray spectrum in all nights. This is true for Nights 1, 3 and 6, where an extrapolation of the X-ray spectrum underpredicts the \g-ray flux. In Nights 2 and 4 the spectra can be connected by a simple extrapolation. This suggests at least for Nights 1, 3 and 6 that the high-energy SED component cannot be described by a single radiation process. This is in line with many FSRQ modelings, where the X-ray domain is reproduced with SSC emission, while the \g\ rays are IC emission scattering jet-external photons.

FSRQs typically host a BLR and a DT. While the BLR luminosity has recently been constrained in \source\ with $L_{\rm BLR}\sim 10^{43}\,$erg/s \citep{Goldoni+24}, the DT parameters are unknown. However, if the emission region were located within the BLR, the optical to UV photons would strongly attenuate the VHE \g-ray emission \citep[e.g.,][]{dermer+12,boettcherels16,meyerscarglebladford19}. A simple estimate can be derived from the fact that \g-\g\ pair production becomes important when the optical depth, $\tau_{\rm BLR}=\sigma_TR_{\rm BLR}n_{\rm BLR}$, becomes larger than unity, with the radius of the BLR, $R_{\rm BLR}$, and the photon density of the BLR given by $n_{\rm BLR} = L_{\rm BLR}/(4\pi c E_{\rm BLR} R_{\rm BLR}^2)$.
Here, $c$ is the speed of light, and $E_{\rm BLR}$ is the energy of a BLR photon. Thus, the optical depth can be estimated as

\begin{align}
    \tau_{\rm BLR} \approx 20 \est{L_{\rm BLR}}{1\E{43}\,\mbox{erg/s}}{} \est{R_{\rm BLR}}{0.1\,\mbox{pc}}{-1} \est{E_{\rm BLR}}{2\,\mbox{eV}}{-1}
    \label{eq:taublr},
\end{align}
which is indeed larger than unity. In Eq.~\eqref{eq:taublr}, the energy of the H$\alpha$ line was used, which has the highest luminosity according to \cite{Goldoni+24}.

Another useful estimate with respect to the absorption in an external photon field comes from the absorption condition $\epsilon_{\g}>2/\epsilon$, where $\epsilon_{\gamma}$ and $\epsilon$ are the \g-ray and soft photon energy normalized to the electron restmass energy, $m_ec^2$, respectively. In turn, observed \g-ray energies, $E_{\g,{\rm obs}}$, are absorbed if

\begin{align}
    E_{\g,{\rm obs}} > 480\,\mbox{GeV}\ \est{\delta}{50}{} \est{\Gamma_b}{50}{-1} \est{T\p}{10^4\,\mbox{K}}{-1},
    \label{eq:egammaobslim}
\end{align}
where $T\p$ is the temperature of the external photon field. 
The relationship between $\delta$ and $\Gamma_b$ implies that their ratio is at most 2.
The high optical depth in the BLR, and the observation of VHE \g\ rays beyond 1\,TeV from \source\ imply that the emission region is located at the edge or beyond the BLR (typical BLR temperature $T\p>10^4$); in line with many other FSRQs \citep[e.g.,][]{costamante+18,hess19,hess+21}. Hence, the IC/DT process is more important than the IC/BLR process.

Using the assumption that electron cooling is dominated by IC scattering of external photons, a constraint can be set on the electron Lorentz factor $\gamma_{KN}$ at which point the IC process enters the Klein-Nishina domain. This happens for $4\gamma_{KN}\epsilon=1$, or

\begin{align}
    \gamma_{KN} = 3\E{3} \est{\Gamma_b}{50}{-1} \est{T\p}{10^4\,\mbox{K}}{-1}.
\end{align}
This suggests that the photons observed at the highest energies have been emitted in the Klein-Nishina regime. In this case, the highest achievable scattered photon energy $\epsilon_s$ equals the maximum electron Lorentz factor $\gamma_{\rm max}$. While it is unknown if the observed maximum energy, $E_{\rm max,obs}\sim 1\,$TeV, corresponds indeed to the maximum energy produced in the emission region, it at least provides a lower limit on $\gamma_{\rm max}$:

\begin{align}
    \gamma_{\rm max}>5\E{4} \est{E_{\rm max,obs}}{1\,\mbox{TeV}}{} \est{\delta}{50}{-1}.
    \label{eq:gammax}
\end{align}

The ratio of the IC peak flux to the synchrotron peak flux is generally referred to as the Compton dominance, $\CD$. In the low state, the Gaia measurement of \source\ along with the archival measurements below $\sim 0.1\,$eV indicate that the synchrotron SED is rather flat, and that $\CD\sim 10$.
While the synchrotron flux is poorly constrained during the flare due to \thestar, it is still clear that $\CD\gg 10$ during the flare, and maybe up to $100$, 
implying a significant increase of $\CD$.
Such a behavior of the $\CD$ along with the moderate variability in the X-ray domain has also been observed in other FSRQs \citep[e.g.,][]{zacharias+17,hess19}.
As the fluxes in the synchtrotron and IC component in the Thomson limit depend linearly on the underlying magnetic field and soft photon energy densities, respectively, the $\CD$ is a direct measure of the ratio of these energy densities. Hence,

\begin{align}
    \CD &= \frac{u_{\rm ext}}{u_B} = \frac{4}{3}\Gamma_b^2 \frac{L\p_{\rm ext}}{4\pi c R^{\prime 2}_{\rm ext}}\, \frac{8\pi}{B^2} \nonumber \\
    \Leftrightarrow \frac{B}{\Gamma_b} &= \sqrt{\frac{8 L\p_{\rm ext}}{3c\, \CD\, R^{\prime 2}_{\rm ext}}} \nonumber \\
    &= 0.01\,\mbox{G}\ \est{\CD}{10}{-1/2} \est{L\p_{\rm ext}}{1\E{44}\,\mbox{erg/s}}{1/2} \est{R\p_{\rm ext}}{1\,\mbox{pc}}{-1},
    \label{eq:Bfield}
\end{align}
with the luminosity, $L\p_{\rm ext}$, and the radius, $R\p_{\rm ext}$, of the external photon field in the host frame assuming that the external photon field fully immerses the jet emission region. On the one hand, given that $\CD$ may be up to $100$ during the flare, nominal BLR parameters ($R\p_{\rm BLR}\sim 0.1\,$pc) and $\Gamma_b\gtrsim 20$ would suggest magnetic fields on the order of $1\,$G, as is typically assumed in FSRQs \citep[e.g.,][]{boettcher+13,hess19,hess23}. On the other hand, as established already, it is more likely that the DT photon field serves as the main IC target. Typical DT radii are taken as $R\p_{\rm DT}\sim 10\,$pc. In order to achieve similar magnetic fields while keeping $\Gamma_b\lesssim 50$ \citep[as inferred from radio observations of blazars,][]{Jorstad+17}, a much larger DT luminosity would be needed. However, the archival IR data provides an upper limit on the DT flux (see the explicit modeling below). Under the assumption that the DT has not changed its characteristics during the flare, it would need an unusually small radius in order to keep the magnetic field in the range $0.1-1\,$G during the flare. With more reasonable DT parameters, the magnetic field must have been relatively low, $B<0.1\,$G, during the flare even with $\Gamma_b=50$.

%%%%%%%%%%%%%%%%%%%%%%%%%%%%%%%%%%%%%%%%%%%%%%%%%%
\subsection{Modeling strategy} \label{sec:modelstrat}
In order to account for the presence of \thestar\ in the optical domain, the following crude strategy is followed. A fit-by-eye of a Planck spectrum (red dotted line in Fig.~\ref{fig:spec_model}; with parameters $T^{\ast}=4700\,$K and distance normalization\footnote{This quantity provides the flux reduction due to the $d^{-2}$ law. It is a fitting value and no actual distance measure is meant to be derived.} $D^{\ast}=3\E{-22}$) to the red asterisks of the Gaia data is done to approximate the stellar spectrum. The parameters should not be considered as real values of \thestar, but merely as roughly representing the de-absorbed Gaia data points. The next step is to include the synchrotron component of the blazar. The sum of the blazar's synchrotron flux and \thestar's flux should then fit the black data points from the \textit{Swift}-UVOT and ATOM data.

Owing to the significant change between the low state and the flaring state, the flaring region is considered to be spatially separated from the low-state region. Hence, a low-state model is included (blue line in Fig.~\ref{fig:spec_model}) providing a baseline flux during the flare and making this effectively a two-zone model. This follows the findings in the FSRQ PKS~1510-089, where a two-zone model has been established \citep{hess23}. Thus, the total spectrum (black line in Fig.~\ref{fig:spec_model}) is the sum of \thestar, the low-state spectrum and the flare model. 

The blazar emission is reproduced using a leptonic model with synchrotron, SSC and IC/DT emission employing the \onehale\ code \citep[version 1.1,][]{zacharias21,zacharias+22}. The input particle distribution to this code is a simple power-law with spectral index $s$ between a minimum and maximum electron Lorentz factor, $\gamma_{\rm min}$ and $\gamma_{\rm max}$, respectively. The time-dependent Fokker-Planck equation is then evolved until a steady-state solution for the particle distribution is reached. For each model, the DT parameters are kept constant with temperature $T_{\rm DT}\p = 1.5\E{3}\,$K, radius $R_{\rm DT}\p = 6.0\E{18}\,$cm, and luminosity $L_{\rm DT}\p = 1.0\E{44}\,$erg/s. 
They are not constrained by observations, but are bound by the archival data points in the infrared domain (cf. the little bump in the blue line at $\log_{10}(E/\mbox{eV})\sim -0.5$ in Fig.~\ref{fig:spec_model}). The jet emission regions (low and flaring state) are both located within the DT but beyond the BLR. The common assumption $\delta=\Gamma_b$ is used. The low state and each night are reproduced as individual steady-states, and their input parameters can be found in Tab.~\ref{tab:modelparam}. The fit to the data is done ``by eye'', and as few parameters as possible are varied from night to night.

Important output parameters of the model include the power in the observer's frame of the important constituents: electrons, protons, magnetic field and radiation. The powers are calculated by

\begin{align}
    P_{i,{\rm obs}} = \pi R \Gamma_b^2 c u_i
    \label{eq:lumgen},
\end{align}
where $u_i$ is the energy density of the constituent. For the proton energy density, one cold proton per electron is assumed. This can be regarded as an upper limit, as a significant fraction of positrons would lower the proton power. The radiation energy density is calculated following \citet[Eq.~(2)]{zachariaswagner16}. The powers are also listed in Tab.~\ref{tab:modelparam}.

%%%%%%%%%%%%%%%%%%%%%%%%%%%%%%%%%%%%%%%%%%%%%%%%%%
\subsection{Results} \label{sec:modelres}
\begin{figure*}
\centering
\includegraphics[width=0.98\textwidth]{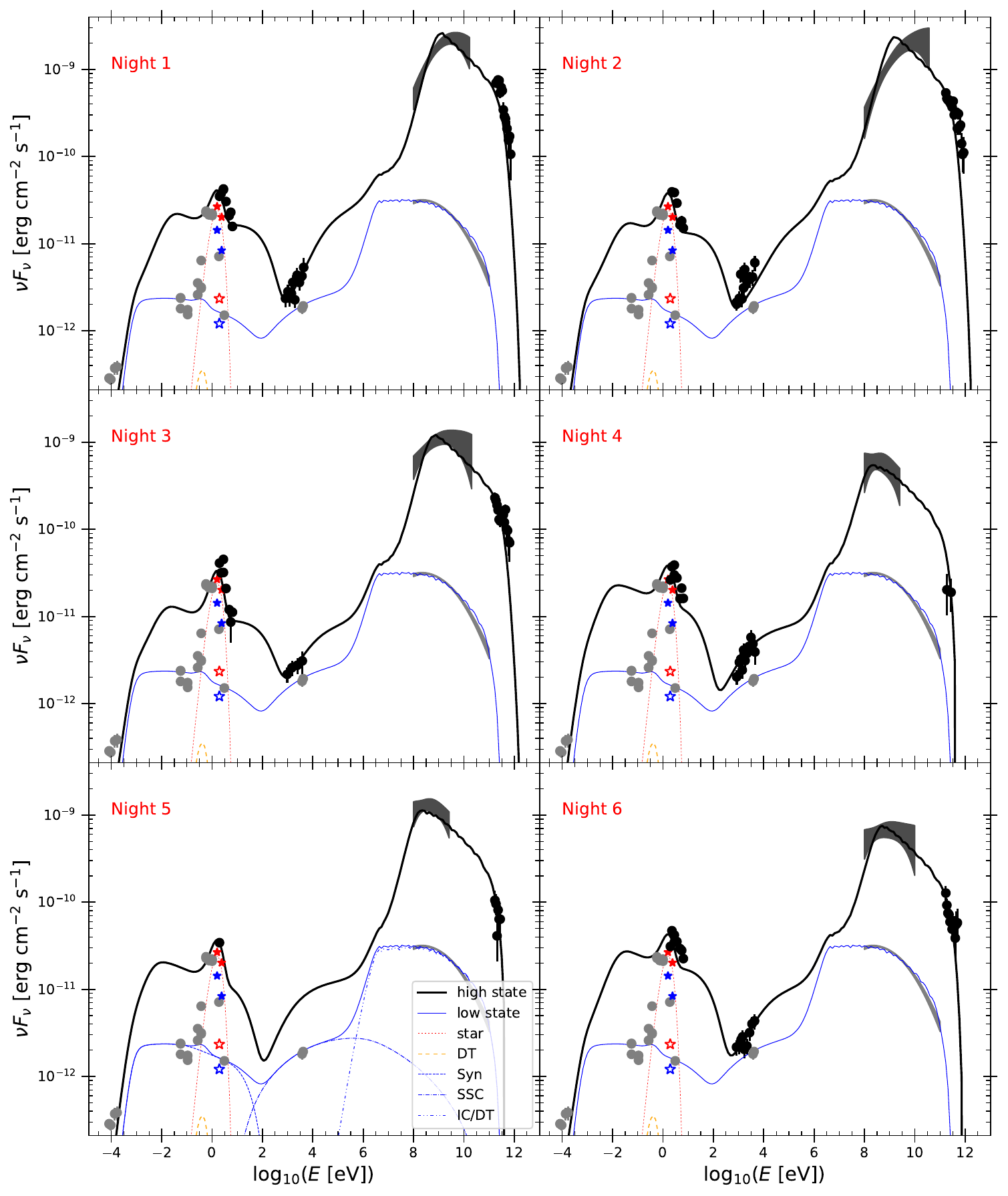}
\caption{MWL spectra for each night with \hess\ observations. The black points and dark butterflies are near-simultaneous with the \hess\ observations. The \hess\ data points are the Mono result corrected for EBL absorption. Gray points mark archival data taken from the NED, while the light gray butterfly is the Fermi-LAT 4FGL spectrum. The blue filled asterisks are the Gaia flux points for \thestar, while the red filled asterisks are the same but corrected for Galactic extinction. The open asterisks show the Gaia flux points for \source\ with the same color coding as for \thestar.
The lines represent the leptonic modeling. The red dotted line marks the fiducial star spectrum, the orange dashed line the DT, the blue solid line the low-state model, and the black line the sum of the low-state model, the star, and the flare model for each night. Exemplary spectral components for the low-state model are given in the panel of Night 5.
}
\label{fig:spec_model}
\end{figure*}
\begin{table*}
\caption{Modeling input parameter definition, symbol and value for the low state and the 6 flare nights as shown in Fig.~\ref{fig:spec_model}. Input parameters are given in the comoving frame. Parameters are only given once if they remain constant during the flare. The output powers listed below the horizontal line are calculated in the observer's frame.
}
\label{tab:modelparam}
\begin{tabular}{lc|c|cccccc}
Definition				        & Symbol 	    		           & Low state & Night 1 & Night 2 & Night 3 & Night 4 & Night 5 & Night 6 \\
\hline
Magnetic field strength	        & $B$ [$10^{-2}\,$G]			   & $6.0$ & $3.0$ & $2.7$ & $3.3$ & $7.0$ & $4.5$ & $6.5$ \\ 
e$^{-}$ injection luminosity	& $L_{\rm inj}$ [$10^{42}\,$erg/s] & $14$ & $4.5$ & $3.5$ & $2.5$ & $1.5$ & $3.2$ & $1.6$ \\ 
Min. e$^{-}$ Lorentz factor	    & $\gamma_{\rm min}$ [$10^2$]	   & $1.0$ & $7.0$ & $9.0$ & $5.0$ & $3.0$ & $3.0$ & $4.5$ \\ 
Max. e$^{-}$ Lorentz factor	    & $\gamma_{\rm max}$ [$10^4$]	   & $5.0$ & $10.0$ & $10.0$ & $10.0$ & $3.0$ & $3.0$ & $5.0$ \\ 
e$^{-}$ spectral index			& $s$				               & $2.9$ & \multicolumn{6}{c}{$2.9$} \\%$2.9$ & $2.9$ & $2.9$ & $2.9$ & $2.9$ & $2.9$ \\ 
Emission region radius			& $R$ [$10^{16}\,$cm]              & $3.0$ & \multicolumn{6}{c}{$2.0$} \\%$2.0$ & $2.0$ & $2.0$ & $2.0$ & $2.0$ & $2.0$ \\ 
Escape time			            & $t_{\rm esc}$ [$R/c$]	           & $1.0$ & \multicolumn{6}{c}{$1.0$} \\%$1.0$ & $1.0$ & $1.0$ & $1.0$ & $1.0$ & $1.0$ \\ 
Doppler factor      	        & $\delta$			               & $20.0$ & \multicolumn{6}{c}{$50.0$} \\%$50.0$ & $50.0$ & $50.0$ & \hline
Electron power             & $P_{e,{\rm obs}}$ [$10^{45}\,$erg/s]   & $3.9$  & $5.4$  & $4.4$ & $3.2$ & $2.4$  & $5.2$ & $2.2$ \\
(Cold) Proton power               & $P_{p,{\rm obs}}$ [$10^{46}\,$erg/s]   & $14.4$ & $3.6$  & $2.6$ & $2.8$ & $3.9$  & $8.2$ & $2.3$ \\
Magnetic power             & $P_{B,{\rm obs}}$ [$10^{42}\,$erg/s]   & $4.9$  & $3.4$  & $2.7$ & $4.1$ & $18.4$ & $7.6$ & $16.8$ \\
Radiation power            & $P_{rad,{\rm obs}}$ [$10^{44}\,$erg/s] & $1.7$  & $10.0$ & $8.9$ & $4.8$ & $2.4$  & $5.0$ & $3.1$ \\
\end{tabular}
\end{table*}
Given the sparsity of archival data, the main goal of the low-state model (blue line in Fig.~\ref{fig:spec_model}) is to reproduce the curvature in the 4FGL spectrum, as well as to not overpredict the few points at lower energies \citep[e.g.,][]{boettcher+13}. Except for the low magnetic field, the parameters for this model are in line with typical FSRQ model parameters. The sharp break at low \g-ray energies 
is due to a $\delta$-function approximation in the inverse-Compton routine of the \onehale\ code. The relative flatness of the low-state synchrotron component is confirmed with the Gaia measurement.

In order to reproduce the flare data, a much higher Doppler factor is used. This has the benefit of keeping the magnetic field more or less in the same range as in the low state despite the significant increase in $\CD$ [cf. Eq~\eqref{eq:Bfield}]. The value of $\delta=50$ is at the high end of observed blazar Doppler factors \citep[e.g.,][]{Jorstad+17}. The emission region radius is put close to the limit set by Eq.~\eqref{eq:Rlim}. Interestingly, the electron spectral index could be kept constant and at the same value as for the low-state model. 
From night to night, the injection luminosity is varied according to the flux changes in the \g-ray spectrum, and the magnetic field is adjusted to match the changes in the optical data. The maximum electron Lorentz factor, $\gamma_{\rm max}$, is varied to follow the maximum energies observed with \hess\ adhering to Eq.~\eqref{eq:gammax}, while not overpredicting the X-ray data due to synchrotron emission. The choices for the minimum electron Lorentz factor, $\gamma_{\rm min}$, require a compromise. In Nights 1 and 2, a higher $\gamma_{\rm min}$ might enable a better match of the \g-ray peak position. However, this would lead to a worse fit at lower \g-ray energies as well as a much reduced SSC flux, which in turn would not reproduce well the X-ray data. The strong variations in $\g_{\rm min}$ (also compared to the low state) reproduce the significant spectral hardening in the HE \g-ray band as discussed in Sec.~\ref{sec:varia} without changing the electron spectral index.

While one should be careful with overinterpreting the evolution in the model parameters -- as a different modeling ansatz may lead to different patterns -- it is still interesting to note the apparent anticorrelation between the magnetic field and the minimum electron Lorentz factor during the first 4 nights, while the injection luminosity continuously decreases. This behavior matches the overall decline in the VHE \g-ray and X-ray light curves, while mirroring the changes in the HE \g-ray flux and index (cf. Fig.~\ref{fig:lc_mwl_zoom}). The lack of \textit{Swift} data in Night 5 makes it impossible to properly constrain the magnetic field, and thus it is not possible to say if the change in the correlation pattern between the parameters in Nights 5 and 6 truly corresponds to the source behavior.

%
%%%%%%%%%%%%%%%%%%%%%%%%%%%%%%%%%%%%%%%%%%%%%%%%%%%%%%%%%%%%%%%%%%%%%%%%%%%%%%%%%%%%%%%%%
\section{Summary \& Discussion} \label{sec:discon}

The blazar \source\ underwent a major \g-ray outburst in March and April 2020 with fluxes changing by about 2 orders of magnitude compared to quiescence. The flare was followed up with \hess\ leading to the VHE \g-ray discovery of the source. This detection resulted in a significant effort by \cite{Goldoni+24} to eventually establish \source\ as an FSRQ at a redshift of $z=0.2621$ making it the second closest VHE-detected FSRQ to Earth after PKS~0736+017 \citep{hess20pks0736}. 

The HE \g-ray flare was marked by four separate flux peaks. Compared to the multi-year average from the 4FGL-DR3 catalog, the spectrum was significantly harder and variable. The \hess\ observations were conducted after the third HE \g-ray peak (flare C). The overall VHE \g-ray flux decrease was interrupted by rising flux trends on time scales of a few hours during the second and third observation night. 
Owing to the significant spectral variability in the HE band, its integrated-flux evolution did not relate well to the overall flux changes in the VHE band.
The X-ray flux was variable within a factor of a few, while the corresponding spectrum remained stable. The optical flux was also variable, however the interpretation of the optical spectra is strongly influenced by \thestar, which cannot be separated from the blazar in observations with ATOM and \textit{Swift}-UVOT.

The flux and spectral variability in the \g-ray component suggest an intricate interplay of time- and energy-dependent acceleration and cooling processes. This is also highlighted by the constraints put on the source parameters derived from the observations. The subsequent modeling was conducted with a sequence of steady states fitted by eye to the six \hess\ observation nights. This resulted in an apparent anticorrelation between the magnetic field and the minimum electron Lorentz factor, even though a different modeling approach may find different parameter patterns. The data strongly suggests that the emission region was located outside of the BLR. The lack of an absorption feature (beyond the EBL) in the \g-ray spectrum as well as the constraints derived from the variability time scales thus indicate that the DT is the more likely seed photon field, which is in-line with previous results on other blazars \citep[see, e.g.,][]{costamante+18,hess19,hess+21,hess23}. As a consequence, it was necessary to choose the magnetic field to be relativley low, $B<0.1\,$G, for both low and high states comprared to other modeling attempts of FSRQs which generally infer $B\sim 1\,$G \citep[e.g.,][]{boettcher+13,zacharias+17,hess19,hess23}.

The jet power emerging from the modeling (cf. Tab.~\ref{tab:modelparam}) is strongly dominated by the particles. This would not change, even if positrons were present in a significant fraction.
With the given parameters, the jet power would be above the Eddington luminosity of a $10^8\,M_{\odot}$ black hole, which is $1.3\E{46}\,$erg/s. A significant fraction of positrons would change this, however, considerably.

The high particle content and low magnetic field values result in a low magnetization of the jet plasma. Under such conditions, particle acceleration is likely mediated through shocks\footnote{The low-magnetization estimate may be a consequence of the assumed isotropy of the particles. Anisotropic particle distributions could lead to higher magnetic field estimates \citep{tavecchiosobacchi20}, a higher magnetization, and thus the possibility of magnetic reconnection powering the flare \citep{sobacchi+23}.} \citep[e.g.,][]{kirkreville10,vanthieghem+20,zechlemoine21}. Furthermore, the possibility to model the low and high state with the same electron spectral index suggests that the shock parameters were generally similar \citep[e.g., the same ratio of the acceleration to escape time scale governing the resulting spectral index;][]{weidingerspanier15}. An additional turbulent component in the magnetic field may have resulted in the significant changes in $\gamma_{\rm min}$ which are the primary reason within the model for the spectral variations in the HE \g-ray domain. Shocks being the main driver of the variability in this FSRQ is in-line with the aforementioned findings on BL Lac objects \citep{liodakis+ixpe22}. 

In summary, the strong flaring activity in \source\ has shown once more that FSRQs produce \g\ rays outside of the region of influence of the BLR. The intricate evolution of the light curves and modeling parameters suggests a turbulent emission region powered likely by shock acceleration.

%
%%%%%%%%%%%%%%%%%%%%%%%%%%%%%%%%%%%%%%%%%%%%%%%%%%%%%%%%%%%%%%%%%%%%%%%%%%%%%%%%%%%%%%%%%
%________________________________________________________________________________________
%
\section{Acknowledgement}
The authors wish to thank the referee for a constructive report that helped to improve the manuscript.
\noindent The \onehale\ code is available upon reasonable request to M.~Zacharias.
This work has made use of data from the European Space Agency (ESA) mission
{\it Gaia} (\url{https://www.cosmos.esa.int/gaia}), processed by the {\it Gaia}
Data Processing and Analysis Consortium (DPAC,
\url{https://www.cosmos.esa.int/web/gaia/dpac/consortium}). Funding for the DPAC
has been provided by national institutions, in particular the institutions
participating in the {\it Gaia} Multilateral Agreement.
This research has made use of the NASA/IPAC Extragalactic Database (NED), which is funded by the National Aeronautics and Space Administration and operated by the California Institute of Technology.
This research has made use of NASA’s Astrophysics Data System.

\section{Author contributions}
% See https://www.sciencedirect.com/journal/journal-of-high-energy-astrophysics/publish/guide-for-authors under "Article structure"
B.~Bi (Data curation, Formal analysis, Writing – original draft),
J.-P.~Lenain (Data curation, Formal analysis, Writing – original draft),
S.~Pita (Data curation, Validation, Writing – original draft),
A.~Wierzcholska (Data curation, Formal analysis, Writing – original draft),
M.~Zacharias (Conceptualization, Methodology, Writing – original draft),
all other members (Writing – review and editing)

\section{Funding sources}
% See https://www.sciencedirect.com/journal/journal-of-high-energy-astrophysics/publish/guide-for-authors under "Article structure"
% HESS acknowledgement version 202306
The support of the Namibian authorities and of the University of
Namibia in facilitating the construction and operation of H.E.S.S.
is gratefully acknowledged, as is the support by the German
Ministry for Education and Research (BMBF), the Max Planck Society,
the Helmholtz Association, the French Ministry of
Higher Education, Research and Innovation, the Centre National de
la Recherche Scientifique (CNRS/IN2P3 and CNRS/INSU), the
Commissariat à l’énergie atomique et aux énergies alternatives
(CEA), the U.K. Science and Technology Facilities Council (STFC),
the Irish Research Council (IRC) and the Science Foundation Ireland
(SFI), the Polish Ministry of Education and Science, agreement no.
2021/WK/06, the South African Department of Science and Innovation and
National Research Foundation, the University of Namibia, the National
Commission on Research, Science \& Technology of Namibia (NCRST),
the Austrian Federal Ministry of Education, Science and Research
and the Austrian Science Fund (FWF), the Australian Research
Council (ARC), the Japan Society for the Promotion of Science, the
University of Amsterdam and the Science Committee of Armenia grant
21AG-1C085. We appreciate the excellent work of the technical
support staff in Berlin, Zeuthen, Heidelberg, Palaiseau, Paris,
Saclay, Tübingen and in Namibia in the construction and operation
of the equipment. This work benefited from services provided by the
H.E.S.S. Virtual Organisation, supported by the national resource
providers of the EGI Federation.
%

%
%%%%%%%%%%%%%%%%%%%%%%%%%%%%%%%%%%%%%%%%%%%%%%%%%%%%%%%%%%%%%%%%%%%%%%%%%%%%%%%%%%%%%%%%%
%________________________________________________________________________________________
%

\appendix
%
%%%%%%%%%%%%%%%%%%%%%%%%%%%%%%%%%%%%%%%%%%%%%%%%%%%%%%%%%%%%%%%%%%%%%%%%%%%%%%%%%%%%%%%%%
\section{VHE \g-ray cross-check analysis} \label{app:xcheck}
\begin{figure}
\centering
\includegraphics[width=0.48\textwidth]{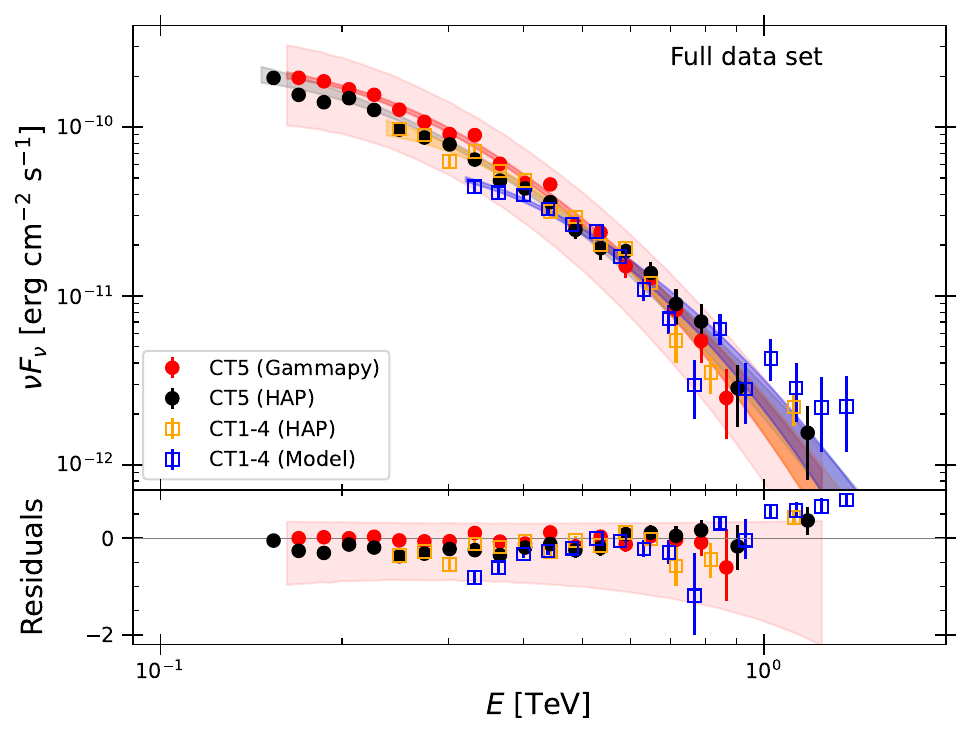}
\caption{
\textit{Top panel:} Total VHE \g-ray spectrum from H.E.S.S. observations. The red spectrum labeled CT5 (Gammapy) is the same as shown in Fig.~\ref{fig:spec_hess_full} including the statistical (dark) and systematics (light) butterflies. The other spectra are from the cross-check analyses as labeled showing only spectral points and the statistical butterflies. The CT1-4 (Model) spectrum is shifted by $6.4\%$ to higher energies to account for the energy bias between the reconstruction chains. 
\textit{Bottom panel:} Residuals of the spectral points with respect to the best fit function, Eq.~\eqref{eq:log-parabola}, of the CT5 (Gammapy) spectrum with parameters given in Tab.~\ref{tab:vhemono}. The light butterfly is the residual of the systematics butterfly of the CT5 (Gammapy) spectrum.
}
\label{fig:spec_hess_xcheck}
\end{figure}

The analysis of the H.E.S.S. data as presented in Sec.~\ref{sec:vhe} is based on events recorded with CT5. These events were calibrated with the HAP analysis chain 
and the spectra have been derived using \texttt{gammapy}. 
The resulting total spectrum is shown again in Fig.~\ref{fig:spec_hess_xcheck} labeled CT5 (Gammapy).
The results have been cross-checked using different reconstruction methods and calibration chains leading to three additional spectra.

The first check was to merely reproduce the CT5 spectrum using H.E.S.S.-internal methods. 
The calibration was also based on the HAP chain.
The resulting spectrum is labeled CT5 (HAP) in Fig.~\ref{fig:spec_hess_xcheck}.

Additional checks were conducted by analysing events recorded with CT1-4 in stereoscopic mode. These events yield a better accuracy in terms of energy and angular resolutions as well as a larger effective area at the overlapping energies, at the cost of a higher energy threshold. Two spectra have been derived employing independent reconstruction chains.

The first CT1-4 analysis was done using the HAP-based ImPACT calibration and analysis chain \citep{parsonshinton14} with \textsc{standard} cuts \citep{hess06}. The corresponding spectrum is labeled CT1-4 (HAP) in Fig.~\ref{fig:spec_hess_xcheck}.

The second CT1-4 analysis reconstructed events using the \textit{Model} analysis \citep{denauroisrolland09} with very loose cuts. These cuts are optimised on Monte Carlo simulations for sources with a soft spectrum such as AGN. Dedicated run-wise simulations \citep{holler20} were produced to incorporate instrument response functions (IRFs) in the spectral analysis matching as closely as possible the actual observation conditions, and thus minimise the systematic errors. 
A comparison between \g-like events from the HAP and Model chains revealed a systematic shift in energy by $6.4\%$. The shifted CT1-4 (Model) spectrum is displayed in Fig.~\ref{fig:spec_hess_xcheck} showing good agreement with the HAP-calibrated spectra.

The good agreement between the various spectra is also indicated by the residuals shown in the bottom panel of Fig.~\ref{fig:spec_hess_xcheck}. The residuals, ${\rm Res}(E)$, are calculated as

\begin{align}
    {\rm Res}(E) = \frac{F(E)-\td{N}(E)/\td{E}}{F(E)}
    \label{eq:app_residuals}
\end{align}
where $F(E)$ denotes the flux point of a spectrum at energy $E$ and $\td{N}(E)/\td{E}$ is the best-fit function of the CT5 (Gammapy) spectrum at the same energy. The best-fit function follows Eq.~\eqref{eq:log-parabola} with the parameters given in Tab.~\ref{tab:vhemono}.

The spectra displayed in Fig.~\ref{fig:spec_hess_xcheck} 
agree very well with each other, and basically all of the corresponding residuals lie within the expected range of the systematics. 
Overall, this indicates that the systematics of the FlashCam camera in CT5, as well as the spectral normalization are well under control.

%
%%%%%%%%%%%%%%%%%%%%%%%%%%%%%%%%%%%%%%%%%%%%%%%%%%%%%%%%%%%%%%%%%%%%%%%%%%%%%%%%%%%%%%%%%
\section{Constraint on the Doppler factor} \label{app:doppler}
A constraint on the Doppler factor, $\delta$, can be derived from internal \g-\g\ absorption following the outline of \cite{dondighisellini95}; see also \cite{boettcher+12}. The optical depth, $\tau_{\g\g}$, as a function of comoving \g-ray photon energy, $E_{\g}$, is

\begin{align}
    \tau_{\g\g}(E_{\g}) = \frac{\sigma_T}{5} \frac{L(E_t)}{4\pi m_ec^3 R}
    \label{eq:app_tau1},
\end{align}
with the Thomson cross-section, $\sigma_T$, the electron rest mass, $m_e$, the speed of light, $c$, and the comoving target photon energy, $E_t$. The intrinsic luminosity of the absorber, $L(E_t)$, can be related to the observed spectral flux, $\nF(\Eto)$, through

\begin{align}
    L(E_t) = \frac{4\pi d_L^2}{1+z} \delta^{-3} \frac{m_e c^2}{\Eto} \nF\left( \Eto \right),
\end{align}
where 

\begin{align}
    \Eto=\frac{\delta^2(m_ec^2)^2}{(1+z)^2 \Ego} 
\end{align} 
and $\Ego$ are the target and \g-ray photon energies in the observer's frame, respectively.
The relation between $\Eto$ and $\Ego$ follows similar steps as those that led to Eq.~\eqref{eq:egammaobslim}. Replacing $\Eto$ leads to

\begin{align}
    L(E_t) = 4\pi d_L^2 \delta^{-5} (1+z) \frac{\Ego}{m_e c^2} \nF\left( \Eto \right),
\end{align}
and hence

\begin{align}
    \tau_{\g\g}(E_{\g}) = \frac{\sigma_T d_L^2}{5c (m_ec^2)^2} \frac{\Ego (1+z)}{\delta^5 R} \nF\left( \Eto \right).
    \label{eq:app_tau2}
\end{align}

Calculating the observed target photon energy, one finds

\begin{align}
    \Eto = 0.16\,\delta^2\est{\Ego}{1\,\mbox{TeV}}{-1}\,\mbox{eV}.
\end{align}
For $\delta=20$, this is in the far-UV regime, while for $\delta=50$ one reaches the soft X-ray domain. Judging from Fig.~\ref{fig:spec_model}, the target photons originate either from the synchrotron component (for lower $\delta$) or from the IC component (for higher $\delta$). This also implies that for lower $\delta$, the $\nF$-flux to be included in Eq.~\eqref{eq:app_tau2} is higher than in case of a higher $\delta$.

If the target photons come from the synchrotron component, the spectral shape in the SED is nearly flat, and we can approximate

\begin{align}
    \nF\left( \Eto \right) \approx (\nF)_0 = \mbox{const.}
\end{align}
Rearranging and using the condition $\tau_{\g\g}(E_{\g})<1$, we obtain the limit on the Doppler factor:

\begin{align}
    \delta &> \left[ \frac{\sigma_T d_L^2}{5c (m_ec^2)^2} \frac{\Ego (1+z)}{R} (\nF)_0 \right]^{1/5} \nonumber \\
    &= 8.9 \est{\Ego}{1\,\mbox{TeV}}{1/5} \est{R}{10^{16}\,\mbox{cm}}{-1/5} \est{(\nF)_0}{2.5\E{-12}\,\frac{\mbox{erg}}{\mbox{cm}^2\,\mbox{s}}}{1/5}
    \label{eq:app_tauopt}.
\end{align}
The example flux used for $(\nF)_0$ resembles the quiescence flux. However, even with an increase of the synchrotron flux by an order of magnitude as probable during the flare, the constraint would only slowly rise to $\sim 15$.

If the target photons are of IC origin in the X-ray domain, the SED can be approximated by a power law:

\begin{align}
    \nF(\Eto) \approx N \left( \frac{\Eto}{E_x} \right)^{\alpha},
\end{align}
where $\alpha=2-\Gamma$. The photon index, $\Gamma$, as well as the normalization, $N$, are given in Tab.~\ref{tab:xrt} for examples during the flare. The normalization is done at energy $E_x=1\,$keV. Replacing again $\Eto$, one finds

\begin{align}
    \nF(\Ego) = N \left( \frac{\delta m_ec^2}{1+z} \right)^{2\alpha} (E_x\Ego)^{-\alpha}.
\end{align}
Plugging this into Eq.~\eqref{eq:app_tau2} results in

\begin{align}
    \delta &> \left[ \frac{\sigma_T d_L^2}{5c} \frac{N\Ego}{R} \left( \frac{m_ec^2}{1+z} \right)^{2-2\Gamma} (E_x\Ego)^{\Gamma-2} \right]^{1/(1+2\Gamma)} \nonumber \\
    &= 0.43 \est{\Ego}{1\,\mbox{TeV}}{\frac{\Gamma-1}{1+2\Gamma}} \est{R}{10^{16}\,\mbox{cm}}{\frac{-1}{1+2\Gamma}} \est{N}{4\E{-4}\,\frac{1}{\mbox{cm}^2\,\mbox{s}\,\mbox{keV}}}{\frac{1}{1+2\Gamma}} 
    \label{eq:app_taux}.
\end{align}
Here, the photon index was set to the long-term average of $\Gamma=1.68$, and $N$ was chosen to represent the low-state flux. As expected, due to the lower X-ray flux compared to the synchrotron flux, the constraint is much less strict than in Eq.~\eqref{eq:app_tauopt}. As the X-ray flux only varies by a factor of a few (at least during this flare), the constraint will not change much. Thus, one ends up in the curious situation that a modeling with a higher $\delta$ (i.e., a higher value of the target photon energy) puts less constraints on it than using a lower one.

%
%%%%%%%%%%%%%%%%%%%%%%%%%%%%%%%%%%%%%%%%%%%%%%%%%%%%%%%%%%%%%%%%%%%%%%%%%%%%%%%%%%%%%%%%%
%________________________________________________________________________________________
%


\bibliographystyle{elsarticle-harv} 
\bibliography{references}


%\begin{thebibliography}{00}
%\bibitem[Lamport(1994)]{lamport94}
%  \textit{\LaTeX: a document preparation system},
%  Addison Wesley, Massachusetts,
%\end{thebibliography}
\end{document}